\documentclass[twocolumn,aps,prl,superscriptaddress]{revtex4-2}
\usepackage{amssymb} 
\usepackage{amsmath,bm} 
\usepackage{graphicx} 
\usepackage[normalem]{ulem} 
\usepackage{multirow} 
\usepackage[colorlinks,linkcolor=blue,urlcolor=blue,citecolor=blue]{hyperref} 
\usepackage{lipsum} 
\usepackage[usenames, dvipsnames]{xcolor} 
\usepackage{tensor} 
\usepackage{isotope} 
\usepackage{amsmath} 
\allowdisplaybreaks[4] 

\renewcommand{\sout}{\bgroup \color{red} \ULdepth=-.5ex \ULset}

\begin{document}

\title{Probing the internal structures of $p\Omega$ and $\Omega\Omega$ with their production at the LHC}
\author{Jie Pu}
\affiliation{College of Physics, Henan Normal University, Xinxiang 453007, China}
\affiliation{Shanghai Research Center for Theoretical Nuclear Physics, NSFC and Fudan University, Shanghai, 200438, China}
\author{Kai-Jia Sun}
\email{Corresponding author: kjsun@fudan.edu.cn}
\affiliation{Key Laboratory of Nuclear Physics and Ion-beam Application (MOE), Institute of Modern Physics,  Fudan University, Shanghai, 200433, China}
\affiliation{Shanghai Research Center for Theoretical Nuclear Physics, NSFC and Fudan University, Shanghai, 200438, China}
\author{Chun-Wang Ma}
\affiliation{College of Physics, Henan Normal University, Xinxiang 453007, China}
\affiliation{Institute of Nuclear Science and Technology, Henan Academy of Science, Zhengzhou, 450046, China}
\affiliation{Shanghai Research Center for Theoretical Nuclear Physics, NSFC and Fudan University, Shanghai, 200438, China}
\author{Lie-Wen Chen}
\email{Corresponding author: lwchen$@$sjtu.edu.cn}
\affiliation{School of Physics and Astronomy, Shanghai Key Laboratory for
Particle Physics and Cosmology, and Key Laboratory for Particle Astrophysics and Cosmology (MOE),
Shanghai Jiao Tong University, Shanghai 200240, China}
\date{\today} 

\begin{abstract}
The strange dibaryons $p\Omega$ ($^5\rm{S}_2$) and $\Omega\Omega$ ($^1\rm{S}_0$) are likely bound, existing either in molecular states like the deuteron or as more exotic compact six-quark states. Here, we investigate the production of these two dibaryons in Pb+Pb collisions at $\sqrt{s_{NN}}$=2.76 TeV at the CERN Large Hadron Collider (LHC) within a covariant coalescence model, which employs a blast-wave-like parametrization for the phase-space configurations of constituent particles at freeze-out. For the molecular states, the $p\Omega$ and $\Omega\Omega$ are produced via $p$-$\Omega$ and $\Omega$-$\Omega$ coalescence, respectively, while for the six-quark states, they are formed through $uudsss$ and $ssssss$ coalescence. We find that the yield ratio $N_{p\Omega}/N_{\Omega}$ and $N_{\Omega\Omega}/N_{\Omega}$ have a distinct centrality dependence between the molecular and multi-quark states, thus offering a promising way for distinguishing the two states.
Our results suggest that the measurements of $p\Omega$ and $\Omega\Omega$ production in relativistic heavy-ion collisions can shed light on their internal structures.
\end{abstract}


\maketitle

\section{Introduction}
\label{introduction}
Exploring dibaryons is a long-standing theoretical and experimental challenge in hadron and nuclear physics~\cite{Clement:2016vnl,ExHIC:2017smd}.
A dibaryon is defined in quantum chromodynamics (QCD) as a six-quark system with baryon number $B=2$ ~\cite{Oka:1988yq,Gal:2015rev}, and the only stable dibaryon observed so far is the deuteron~\cite{PhysRev.39.164}. 
The $H$ dibaryon, a hypothetical bound state with strangeness $S=-2$ and comprising  $uuddss$ quarks with $J^\pi=0^+$ and $I = 0$, was first predicted by Jaffe~\cite{Jaffe:1976yi} within a bag-model approach.
Other interesting candidates for strange dibaryons include the $N\Omega$ ($uudsss$ or $uddsss$) with $J^\pi=2^+$ and $I = 1/2$, and the $\Omega\Omega$ ($ssssss$ ) with $J^\pi=0^+$ and $I = 0$ ~\cite{Goldman:1987ma,HALQCD:2014okw,Morita:2016auo}.
Within various quark models, the $N\Omega$ dibaryon has been suggested to be a bound state~\cite{Wang:1995bg,Huang:2015yza} or quasibound state~\cite{Sekihara:2018tsb}. In addition, the $\Omega\Omega$ dibaryon was predicted to be a bound state by a chiral quark model~\cite{Zhang:1997ny,Zhang:2000sv,Li:2000cb,Dai:2006gs,Huang:2019hmq}, while a weak repulsive $\Omega$-$\Omega$ interaction was suggested by other models~\cite{Wang:1995bg,Buchoff:2012ja}.
Recent lattice QCD studies using the time-dependent HAL QCD method have shown the $N\Omega(^5S_2)$ potential to be attractive at all distances, producing a quasi-bound state near unitarity~\cite{HALQCD:2018qyu,Garcilazo:2018gkb,Chen:2021hxs}. The di-Omega ($\Omega\Omega$)($^1S_0$) was also thoroughly examined using lattice QCD data analyzed with the HAL QCD method and the results show that the $\Omega\Omega$ system has an overall attraction and is located near the unitary regime~\cite{Gongyo:2017fjb}. The $\Omega\Omega$, being the most strange dibaryon with $S=-6$, is of great interest in QCD~\cite{Gongyo:2017fjb,Chen:2019vdh}.

Experimentally, high-energy nuclear collisions provide a unique tool to extract information on nucleon-hyperon and hyperon-hyperon interactions via measuring the hadron-pair correlation functions~\cite{STAR:2014dcy,STAR:2015kha,Chen:2018tnh}.
The STAR Collaboration has measured the $p$-$\Omega$ correlation function~\cite{STAR:2018uho} in Au+Au collisions at $\sqrt{s_{NN}}= 200$~GeV and the results suggest a bound state of $p$-$\Omega$ with a binding energy of $27$~MeV.
The ALICE Collaboration's measurements in p+p collisions $\sqrt{s_{NN}}= 13$~TeV indicated that the correlation functions predicted by lattice QCD underestimate the data, warranting further investigations to conclusively ascertain the existence of this bound state~\cite{ALICE:2020mfd}. 

Exploration of the production of $p\Omega$ and $\Omega\Omega$ dibaryons in high-energy nuclear collisions may offer insights into their nature~\cite{Lin:2021mdn,ExHIC:2010gcb,ExHIC:2011say,Zhang:2020dma}.
The production of $p\Omega$ and $\Omega\Omega$ dibaryons in high energy heavy-ion collisions has been investigated within the coalescence model by assuming the $p\Omega$ and $\Omega\Omega$ dibaryons are bound molecular states of $p$-$\Omega$ ($^5\rm{S}_2$) and $\Omega$-$\Omega$ ($^1\rm{S}_0$), respectively.
For the $p\Omega$ and $\Omega\Omega$ dibaryons, the Pauli exclusion principle does not operate among their valence quarks, and these dibaryons can in principle be compact six-quark states~\cite{Oka:1988yq,Gal:2015rev,Chen:2010zzi}.
In high-energy heavy-ion collisions, the coalescence production of clusters generally depends on the internal structures of the hadrons~\cite{ExHIC:2010gcb,ExHIC:2011say,Cho:2013rpa,Abreu:2016qci,Wu:2020zbx}.
Therefore, it is interesting to probe the internal structure effects on the production of $p\Omega$ ($^5S_2$) and $\Omega\Omega$ ($^1S_0$) in these collisions by treating them as either molecular or six-quark states. This is our main motivation of this work. 

In the present study, we compute the production of the $p\Omega$ and $\Omega\Omega$ exotic states in central Pb+Pb collisions at $\sqrt{s_{NN}}$=2.76 TeV using a covariant coalescence model with a blast-wave-like parametrization for the phase-space configurations of constituent particles at freeze-out. Specifically, we calculate the yields of $p\Omega$ and $\Omega\Omega$ through hadron coalescence in the case of molecular states and through quark coalescence for six-quark states. We find a distinct difference in the centrality dependence of the yield ratios (scaled by the corresponding yield of $\Omega$) between the molecular and multi-quark states, providing a conceivable tool to differentiate between these two structures. 

\section{Covariant Coalescence Model}
The main feature of the coalescence model~\cite{Butler:1961pr,Sato:1981ez,Csernai:1986qf} is that the coalescence probability depends on the details of the phase space configurations of the constituent particles at freeze-out as well as the statistical weight and wave function of the coalesced cluster. At this point, we would like to mention that the details of cluster wavefunction are of no relevance in the thermal model~\cite{Andronic:2010qu,Cleymans:1990mn,Steinheimer:2012tb,Cleymans:2011pe,Braun-Munzinger:1994zkz,Braun-Munzinger:2007edi} for cluster production.
For particle production at mid-rapidity in central Pb+Pb collisions at $\sqrt{s_{NN}}=2.76$~TeV at LHC considered here, we assume a longitudinal boost-invariant expansion for the constituent particles which are emitted from a
freezeout hypersurface $\Sigma^\mu$, and the Lorentz invariant one-particle
momentum distribution is then given by
\begin{eqnarray}
E\frac{\text{d}^3 N}{\text{d}^3 p}=\frac{\text{d}^3 N}{p_T \text{d} p_T \text{d} \phi_p \text{d} y } =  \int \text{d}^4 x S(x,p),
\end{eqnarray}
where $p^\mu$ is the
four-momentum of the emitted particle, $\phi_p$ is the azimuthal direction of the emitted particle, $p_T = \sqrt{p_x^2+p_y^2}$ is the transverse momentum and
$S(x,p)$ is the emission function which reads
\begin{eqnarray}
S(x,p)\text{d}^4x = m_T \text{cosh} (\eta-y)f(x,p)J(\tau)\text{d}\tau \text{d}\eta r \text{d} r \text{d} \phi_s.
\end{eqnarray}
In the above expressions, we use longitudinal proper time $\tau = \sqrt{t^2-z^2}$,
space-time rapidity $\eta = \frac{1}{2} \text{ln}\frac{t-z}{t+z}$, cylindrical
coordinates ($r$, $\phi_s$), rapidity $y=\frac{1}{2}\ln (\frac{E+p_z}{E-p_z})$,
transverse momentum ($p_T,\phi_p$), and transverse mass $m_T=\sqrt{m^2+p_T^2}$. In the relativistic heavy-ion collisions, the statistical distribution function $f(x,p)$ is given
by
\begin{eqnarray}
f(x,p)=g(2\pi)^{-3}[\exp(p^{\mu}u_{\mu}/kT)/\xi \pm 1]^{-1}
\end{eqnarray}
with $g$ being spin degeneracy factor, $\xi$ the fugacity, $u_{\mu}$ the
four-velocity of a fluid element in the fireball, and $T$ the local temperature.
The detailed information can be found in Refs.~\cite{Sun:2016rev,Sun:2015jta}.
The freeze-out process generally occurs gradually over a period of time, and here we assume the freeze-out time follows a Gaussian distribution~\cite{Retiere:2003kf}
\begin{eqnarray}
J(\tau)=\frac{1}{\Delta \tau \sqrt{2\pi}}\exp[-\frac{(\tau-\tau_0)^2}{2(\Delta \tau)^2}]
\end{eqnarray}
with a mean value $\tau_0$ and a dispersion $\Delta \tau$.
The transverse rapidity distribution of the fluid element in the fireball is
parametrized as $\rho=\rho_0 r/R_0$ with $\rho_0$ being the maximum transverse
rapidity and $R_0$ the transverse radius of the fireball.
The phase-space freeze-out configuration of the constituent particles
are thus determined by six parameters, i.e., $T$, $\rho_0$, $R_0$, $\tau_0$,
$\Delta \tau$ and $\xi$.

The cluster production probability can be calculated through
the overlap of the cluster Wigner function with the constituent
particle phase-space distribution at freeze-out. If $M$ particles
are coalesced into a cluster, the invariant
momentum distribution of the cluster can be obtained as
\begin{eqnarray}
E\frac{\text{d}^3N_c}{\text{d}^3P}&=&Eg_c\int  \bigg(\prod_{i=1}^{M} \frac{\text{d}^3p_i }{E_i}\text{d}^4x_iS(x_i,p_i)\bigg)\times \notag \\
&&\rho_c^W(x_1,...,x_M;p_1,...,p_M)\delta^3(\mathbf{P}-\sum_{i=1}^M\mathbf{p_i}),  \notag \\
\label{Eq:Coal}
\end{eqnarray}
where $N_c$ is the cluster multiplicity, $E$ ($\mathbf{P}$) is its
energy (momentum), $g_c$ is the coalescence factor~\cite{Sato:1981ez} which can be expressed as $g_c=\frac{2S+1}{2^M3^M}$ including the spin and color degrees of freedom in the quark coalescence, but $g_c=\frac{2J+1}{(2j_1+1)(2j_2+1)...(2j_M+1)}$ in the hadron coalescence including the spin degrees of freedom, where $j_M$ is spin of particle, and $J$ is the spin of the cluster.
$\rho_c^W$ is the Wigner function and $\delta$-function is adopted to ensure momentum conservation.
In this study, the harmonic oscillator wave functions are assumed and
the cluster Wigner function is expressed as
\begin{eqnarray}
&&\rho_c^W(x_1,...,x_M;p_1,...,p_M) \notag\\
&&\quad= \rho ^{W}(q_{1},\cdot \cdot \cdot ,q_{M-1},k_1,\cdot
\cdot \cdot ,k_{M-1}) \notag\\
&&\quad= 8^{M-1}\exp [-\sum_{i=1}^{M-1}(q_{i}^{2}/\sigma _{i}^{2}+\sigma_{i}^{2}k_i^{2})],
\end{eqnarray}
where $\mu_{i-1}= \frac{i}{i-1} \frac{m_i \sum_{k=1}^{i-1}m_k}{\sum_{k=1}^{k=i}m_k},(i\geq2)$ is the reduced mass in the center-of-mass frame, $\sigma_i^2 = (\mu_{i} w)^{-1}(1\leq i\leq M-1)$, and $w$ is the harmonic oscillator frequency. The detailed information about the  coordinate transformation from $(x_1,...,x_M),(p_1,...,p_M)$ to relative coordinates $(q_1,...,q_{M-1}),(k_1,...,k_{M-1})$   can be found in  Ref.~\cite{Sun:2015jta}.
If the masses of constituent particles are equal, we then have $\mu_i=m$. The mean-square radius is given by
\begin{eqnarray}
\langle r^2_M \rangle = \frac{3}{2M w}[\sum_{i=1}^M \frac{1}{m_i} -\frac{M}{\sum_{i=1}^M m_i}].
\label{rm}
\end{eqnarray}
The integral~(\ref{Eq:Coal}) can be directly calculated  through multi-dimensional numerical integration by the Monte-Carlo method~\cite{Lepage:1977sw,Sun:2015jta}. It should be emphasized that since
the constituent particles may have different freeze-out time, in the numerical calculation, the
particles that freeze out earlier are allowed to propagate freely
until the time when the last particle in the cluster freezes out in order
to make the coalescence at equal time~\cite{Sun:2015jta,Mattiello:1996gq,Chen:2006vc}.

\section{results and discussions}
\subsection{Molecular states from hadron coalescence}
We first consider the $p\Omega$ and $\Omega\Omega$ dibaryons as molecular states, and their production in central Pb+Pb collisions at $\sqrt{s_{NN}}=2.76$ TeV can be described by $p$-$\Omega$ and $\Omega$-$\Omega$ coalescence, respectively, in the covariant coalescence model.
The freeze-out configurations of nucleons (denoted FOPb-$p$) and $\Omega$ (denoted FOPb-$\Omega$) for central Pb+Pb collisions at $\sqrt{s_{NN}}=2.76$ TeV have been obtained by fitting the experimental spectra~\cite{ALICE:2012ovd,ALICE:2013cdo,ALICE:2014jbq,ALICE:2013xmt} of protons, $\Omega$, deuterons, and $^3$He~\cite{Sun:2016rev}.
The parameter values of FOPb-$p$ and FOPb-$\Omega$ are summarized in Tab.~\ref{ParamHadron}, and the details can be found in Refs.\cite{Sun:2015jta, Sun:2015ulc,Sun:2018mqq}.  The freeze-out temperature $T$ of nucleons \cite{Sun:2018mqq} can be extracted from measured charged particle spectra expressed as:
\begin{eqnarray}
T=T_0+T_1 \left[ 1+(q-1)\times \frac{\text{d} N_{ch}/\text{d} \eta}{M} \right] ^{-\frac{1}{q-1}},
\label{Eq:TK}
\end{eqnarray}
by using the four parameters $T_0=80.6\pm31.0$ MeV, $T_1=83.0\pm46.9$ MeV, $q=3.33\pm3.25$, and $M=67.3\pm76.3$ after taking into account the errors. The values of $T$ in different centralities are shown in Tab.~\ref{ParamHadron} (FOPb-$p$). 
According to Ref.~\cite{Sun:2018mqq}, the values of $R_0$ and $\tau_0$ of nucleons in different centralities can be calculated by using the equal source volume $(2\pi)^{1.5}R^3=\pi R_0^2\tau_0$~\cite{STAR:2014shf} and the fixed ratio $R_0/\tau_0=19.7/15.5$, where $R_0$ and $\tau_0$ are the same as ``FOPb-$N$" in Ref.~\cite{Sun:2015ulc}.  
The source radius $R$ can be calculated from
\begin{eqnarray}
R=\frac{(3N_n)^{1/3}}{[4C_1(mT)^{3/2}]^{1/3}},
\label{Eq:R}
\end{eqnarray}
where $N_n$ is the neutron number which is the same as the proton number~\cite{ALICE:2013mez} in collisions at the LHC energies, the factor $C_1=4.0 \times 10^{-3}$ is obtained from the yield ratio d/p \cite{Sun:2018mqq}. The $R_0$ and $\tau_0$ in different centralities are shown in Tab.~\ref{ParamHadron} (FOPb-$p$).     
With the values $R_0$, $\tau_0$ and the temperature $T$, the other parameters can be obtained by fitting the $p$ spectra~\cite{ALICE:2013mez} in the corresponding centrality and shown in Tab.~\ref{ParamHadron} (FOPb-$p$).
Similar to ``FOPb-p", the $T$, $R_0$ and $\tau_0$ of FOPb-$\Omega$ have same values with FOPb-$p$, then the other parameters of FOPb-$\Omega$ can be obtained by fitting the $\Omega$ spectra \cite{ALICE:2013xmt} in the corresponding centrality and shown in Tab. \ref{ParamHadron} (FOPb-$\Omega$).

\begin{figure}
\includegraphics[scale=0.43]{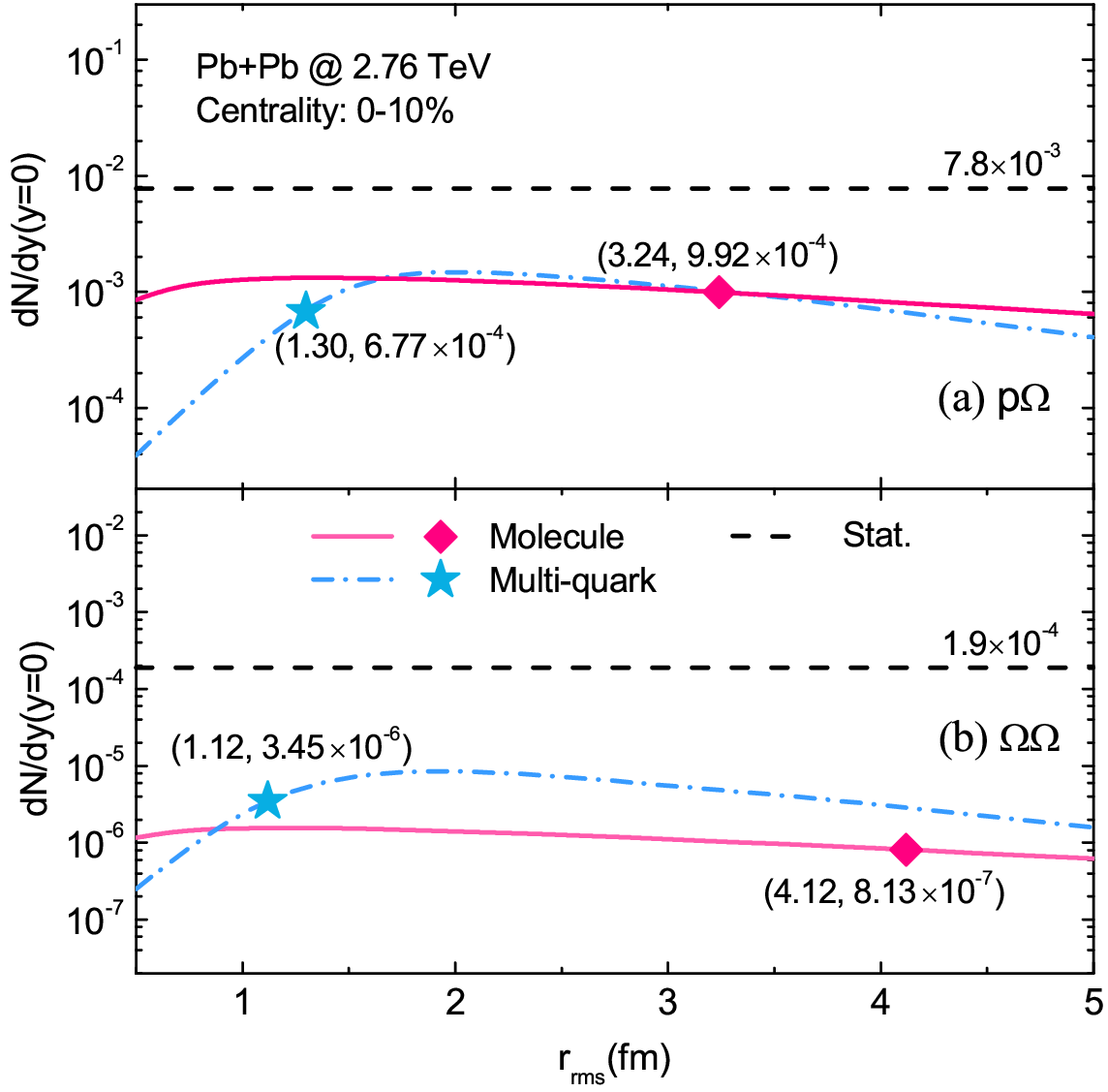}
\caption{\protect\small  The yields of $p\Omega$ (top) and $\Omega\Omega$ (bottom) versus root-mean-square radii through hadron coalescence and quark coalescence by using the parameters of Table \ref{ParamHadron} and \ref{Paramquark} in centrality $0-10\%$. The black dashed lines are results from the statistical model~\cite{ExHIC:2017smd}.} 
\label{FigQuarkLHC}
\end{figure}

The cluster yield also depends on the cluster size~\cite{Sun:2015ulc,Zhang:2020dwn, Cheng:2023nst}, we thus calculate the yields of $p\Omega$ and $\Omega\Omega$ as a function of their root-mean-square radii ($r_{\rm rms}$).
Fig.~\ref{FigQuarkLHC} shows the $p_T$-integrated yield in the mid-rapidity region ($-0.5\leq y\leq0.5$) versus $r_{\rm rms}$ in the range of $0.5 \sim 5$ fm through hadron coalescence (solid lines) in central collision parameters. 
For FOPb-p and FOPb-$\Omega$ in centrality $0$-$10\%$, the yield (d$N/$d$y$) of $p\Omega$ ranges from $ 6.41\times 10^{-4}$ to $1.32 \times 10^{-3}$, and the yield of $\Omega\Omega$ ranges from $6.25\times10^{-7}$ to $1.56\times10^{-6}$.

In lattice QCD calculations, the binding energy of $N\Omega$ ($B_{N\Omega}$) is 1.54 MeV and 2.46 MeV with and without taking into account the Coulomb attraction~\cite{HALQCD:2018qyu}, and the corresponding root mean square distance is $\sqrt{\langle r_M^2 \rangle}= 3.77$ fm and $\sqrt{\langle r_M^2 \rangle}= 3.24$ fm.  The binding energy of $\Omega\Omega$ is $B_{\Omega\Omega}=1.6$ MeV or $B_{\Omega\Omega}=0.7$ MeV with/without Coulomb attraction~\cite{Gongyo:2017fjb} and the corresponding root mean square distance is about $\sqrt{\langle r_M^2 \rangle}= 3.28$ fm and $\sqrt{\langle r_M^2 \rangle}= 4.12$ fm. So we assume the binding energy of $p\Omega$ is 2.46 MeV, the binding energy of $\Omega\Omega$ is 0.7 MeV. The corresponding $r_{\rm rms}$ of $p\Omega$ and $\Omega\Omega$ are 3.24 fm and 4.12 fm~\cite{ExHIC:2011say} which are shown by diamonds in the solid lines in Fig.\ref{FigQuarkLHC}. The yields of $p\Omega$ and $\Omega\Omega$ are $9.92\times10^{-4}$ at $r_{\rm rms}$ = 3.24 fm and $8.13\times10^{-7}$ at $r_{\rm rms} = 4.12$ fm, respectively.
\begin{table}
\caption{Parameters of the blast-wave-like analytical parametrization for the phase-space configurations at freeze-out for (anti-)nucleon (FOPb-$p$)~\cite{Sun:2015jta}, (anti-)$\Omega$ (FOPb-$\Omega$)~\cite{Sun:2015ulc} for Pb+Pb central collisions at $\protect\sqrt{s_{NN}}=2.76$~TeV in different centralities. }
\begin{tabular}{cccccccccccccccccc}
        \hline \hline
                 FOPb-$p$ \\
      \hline  
      Centrality & T(MeV) & $\rho_0$ & $R_0$(fm) & $\tau_0$(fm/c)& $\Delta \tau$  & $\xi_H$ \\
         \hline
      0-10\% & 95.94  & 1.28 & 21.92 & 17.25 & 1.0 & 27.24   \\
      10-20\% & 98.77  & 1.25 & 18.98 & 14.93 & 1.0 & 20.87 \\
      20-40\% & 103.74  & 1.21 & 15.36 & 12.09 & 1.0 & 13.42   \\
      40-60\% & 108.30  & 1.12 & 11.12 & 8.75 & 1.0 & 9.37   \\
      60-80\% & 114.42  & 1.01 & 7.15 & 5.62 & 1.0 & 6.01 \\
      \hline \hline
      FOPb-$\Omega$ \\
      \hline
      Centrality & T(MeV) & $\rho_0$ & $R_0$(fm) & $\tau_0$(fm/c)& $\Delta \tau$  & $\xi_H$ \\
      0-10\% & 95.94  & 1.08 & 21.92 & 17.25 & 1.0 & 454.40  \\
      10-20\% & 98.77  & 1.10 & 18.98 &  14.93& 1.0 & 265.48 \\
      20-40\% & 103.74  & 1.04 & 15.36 & 12.09 & 1.0 & 123.51 \\
      40-60\% & 108.30  & 0.97 & 11.12 & 8.75 & 1.0 & 57.25   \\
      60-80\% & 114.42  & 0.95 & 7.15 & 5.62 & 1.0 & 13.95 \\
        \hline  \hline
\end{tabular}
\label{ParamHadron}
\end{table}

\begin{table}
\caption{Parameters of the blast-wave-like analytical parametrization for the phase-space configurations at freeze-out for light quarks (FOPb-Q)~\cite{Sun:2016rev} for Pb+Pb collisions at $\protect\sqrt{s_{NN}}=2.76$~TeV in different centralities. }
\begin{tabular}{ccccccccc}
        \hline  \hline
        &Centrality & T(MeV) & $\rho_0$ & $R_0$(fm) & $\tau_0$(fm/c)& $\Delta \tau$  & $\xi_u$ & $\xi_s$  \\
         \hline
        &0-10\%  & 154  & 1.08 & 13.6 & 11.0 & 1.3 & 1.02  &0.89  \\
        &10-20\%  & 154  & 1.08 & 12.0 & 9.7 & 1.3 & 1.02  &0.89  \\
        &20-40\%  & 154  & 1.08 & 9.9 & 8.0 & 1.3 & 1.02  &0.89  \\
        &40-60\%  & 157  & 1.03 & 7.3 & 5.94 & 1.3 & 1.02  &0.89  \\
        &60-80\%  & 160  & 0.95 & 4.8 & 3.9 & 1.3 & 1.13  &0.85  \\
        \hline  \hline
\end{tabular}
\label{Paramquark}
\end{table}

\subsection{Six-quark states from quark coalescence}

We now model $p\Omega$ and $\Omega\Omega$ as compact six-quark states, where $p\Omega$ is a $uddsss$ bound state, and $\Omega\Omega$ is a $ssssss$ bound state. Their yields are calculated using the quark coalescence model. By fitting the measured spectra of $p(udd)$ and $\Omega^-(sss)$, as detailed in our previous work~\cite{Sun:2016rev}, we obtain phase-space distributions of $u, d, s$ quarks and their anti-partners at freeze-out. We assume $u$ and $d$ quarks have a mass of 300 MeV, while the $s$-quark mass is 500 MeV. The freeze-out parameters are listed in Tab.~\ref{Paramquark} (FOPb-Q). For instance, in the $0$-$10\%$ centrality, $T=154$ MeV, $\rho_0=1.08$, $R_0=13.6$ fm, $\tau_0=11.0$ fm/$c$, $\Delta\tau_0=1.3$ fm/$c$, $\xi_u=1.02$, and $\xi_s=0.89$ are the temperature, transverse flow rapidity, transverse radius, mean longitudinal proper freeze-out time, proper time dispersion, and fugacities for $u$ and $s$ quarks, respectively (refer to Ref.~\cite{Sun:2016rev} for more details). Using    the quark freeze-out configuration FOPb-Q, we calculate the yields of $p\Omega$ and $\Omega\Omega$ through six-quark coalescence.

The $r_{\rm rms}$ dependence of the yields of $p\Omega$ and $\Omega\Omega$ in $0$-$10\%$ centrality are depicted as dash dot lines in Fig.\ref{FigQuarkLHC}. For quark coalescence, the yield of $p\Omega$ varies from $3.89 \times 10^{-5}$ to $1.54 \times 10^{-3}$, and for $\Omega\Omega$, from $2.49\times10^{-7}$ to $8.80\times10^{-6}$. We assume the harmonic oscillator frequencies for $p\Omega$ and $\Omega\Omega$ are $\omega_s=78$ MeV~\cite{Sun:2016rev}. Consequently, the calculated $r_{\rm rms}$ values for $p\Omega$ and $\Omega\Omega$ are 1.30 fm and 1.12 fm, respectively, indicated by stars in Fig.\ref{FigQuarkLHC}, with corresponding yields of $6.77\times10^{-4}$ and $3.45\times10^{-6}$.
 
In Fig.~\ref{FigQuarkLHC}, for $p\Omega$, it is seen that its estimated yield from hadron coalescence is 1.46 times of that from quark coalescence. For smaller $r_{\rm rms}$, the yield from hadron coalescence is about one order larger than that from quark coalescence.  This is attributed to the significant contribution from strong decays to the proton number, thereby enhancing $p\Omega$ yields through $p$-$\Omega$ coalescence.  While results from ExHIC collaboration also account for the effect of resonance decay, the yield of the molecular state is significantly enhanced but does not exceed that of the multi-quark state~\cite{ExHIC:2011say,ExHIC:2017smd}. For $\Omega\Omega$, the strong decays rarely contribute and the predicted yields  by quark coalescence are higher than those from hadron coalescence at  $r_{\rm rms}>0.87$ fm.  This is consistent with results from ExHIC collaboration~\cite{ExHIC:2011say,ExHIC:2017smd}. 
In the results from both ExHIC collaboration and the present study, the yields of $p\Omega$ and $\Omega\Omega$ are larger in the statistical model than the coalescence model. Since the yields of $p\Omega$ and $\Omega\Omega$ in central heavy-ion collisions depends on not only their internal structure being  molecular states or  compact multi-quark states but also the implementation of coalescence model, deciphering their structure from merely their yields in central heavy-ion collisions is challenging.

\begin{figure}[!t]
\includegraphics[scale=0.35]{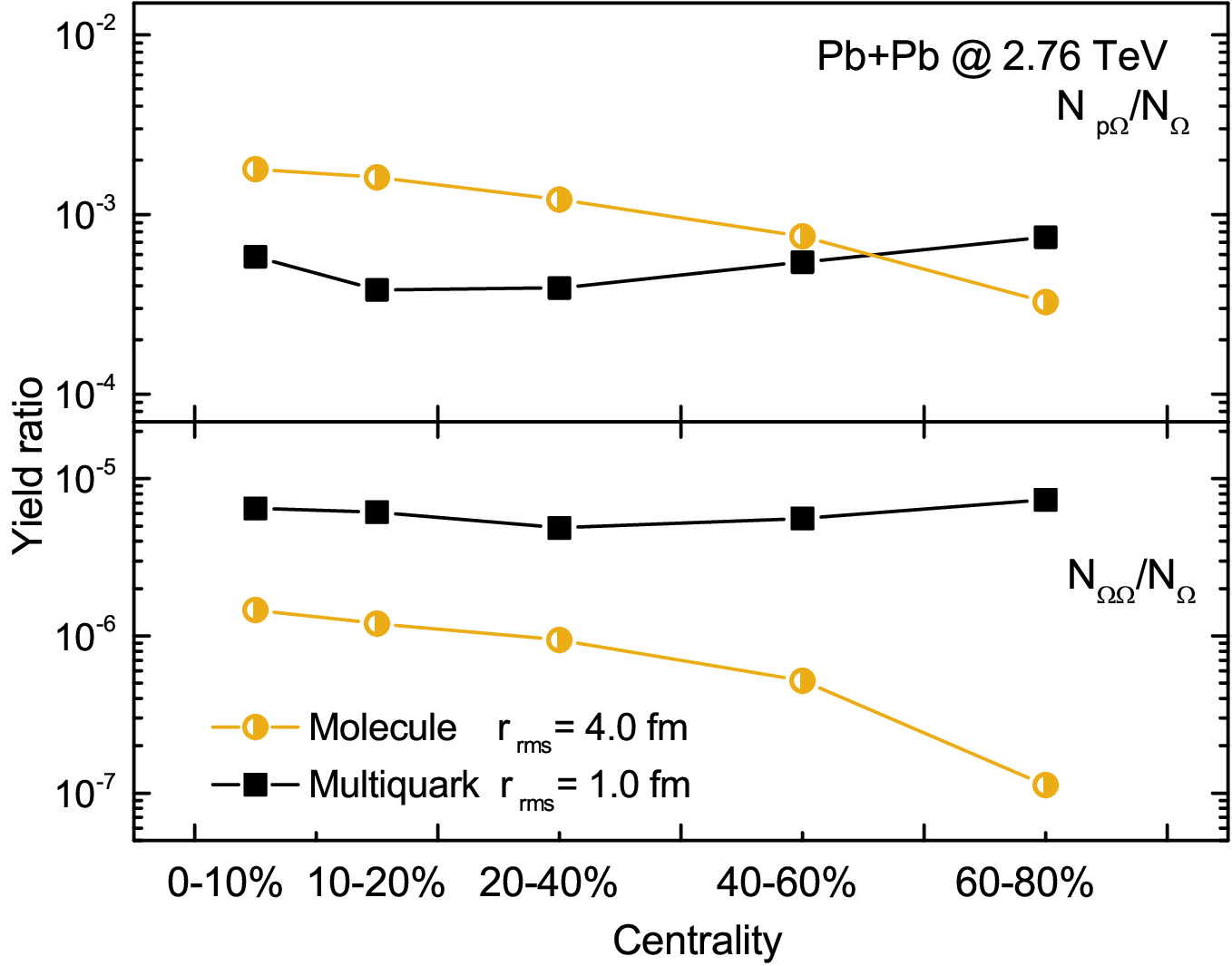}
\caption{\protect\small  The yields ratio of $N_{p\Omega}/N_{\Omega}$ (top) and $N_{\Omega\Omega}/N_{\Omega}$ (bottom) versus centrality. The yields of $p\Omega$ and $\Omega\Omega$ are calculated through hadron coalescence and quark coalescence by using the parameters of Table \ref{ParamHadron} and \ref{Paramquark} with the fixed root-mean-square radii. The yields of $\Omega$ comes form experiment data \cite{ALICE:2012ovd,ALICE:2013cdo,ALICE:2014jbq,ALICE:2013xmt}. }
\label{Figradio-test}
\end{figure}

Figure~\ref{Figradio-test} shows the centrality dependence of yield ratios $N_{p\Omega}/N_{\Omega}$ (top) and $N_{\Omega\Omega}/N_{\Omega}$ (bottom), with fixed $r_{\rm rms} = 1.0$ fm for quark coalescence and $r_{\rm rms} = 4.0$ fm for hadron coalescence. In Fig.~\ref{Figradio-test} (a), the yellow circles represent the $p\Omega$ yield ratio from hadron coalescence to $\Omega$ from experimental data~\cite{ALICE:2013xmt}, showing a rapid decrease from central to peripheral collisions. The black squares show the yield ratio of $p\Omega$ from 6-quark coalescence to $\Omega$ from experimental data, with a mild change from central to peripheral collisions. Fig.~\ref{Figradio-test} (b) displays the $N_{\Omega\Omega}/N_{\Omega}$ ratio, indicating a rapid drop from central to peripheral collisions in hadron coalescence but only a mild change in the multi-quark coalescence. These different centrality dependencies  provide a promising tool  for distinguishing molecular and multi-quark states.

\begin{figure}[!t]
\includegraphics[scale=0.35]{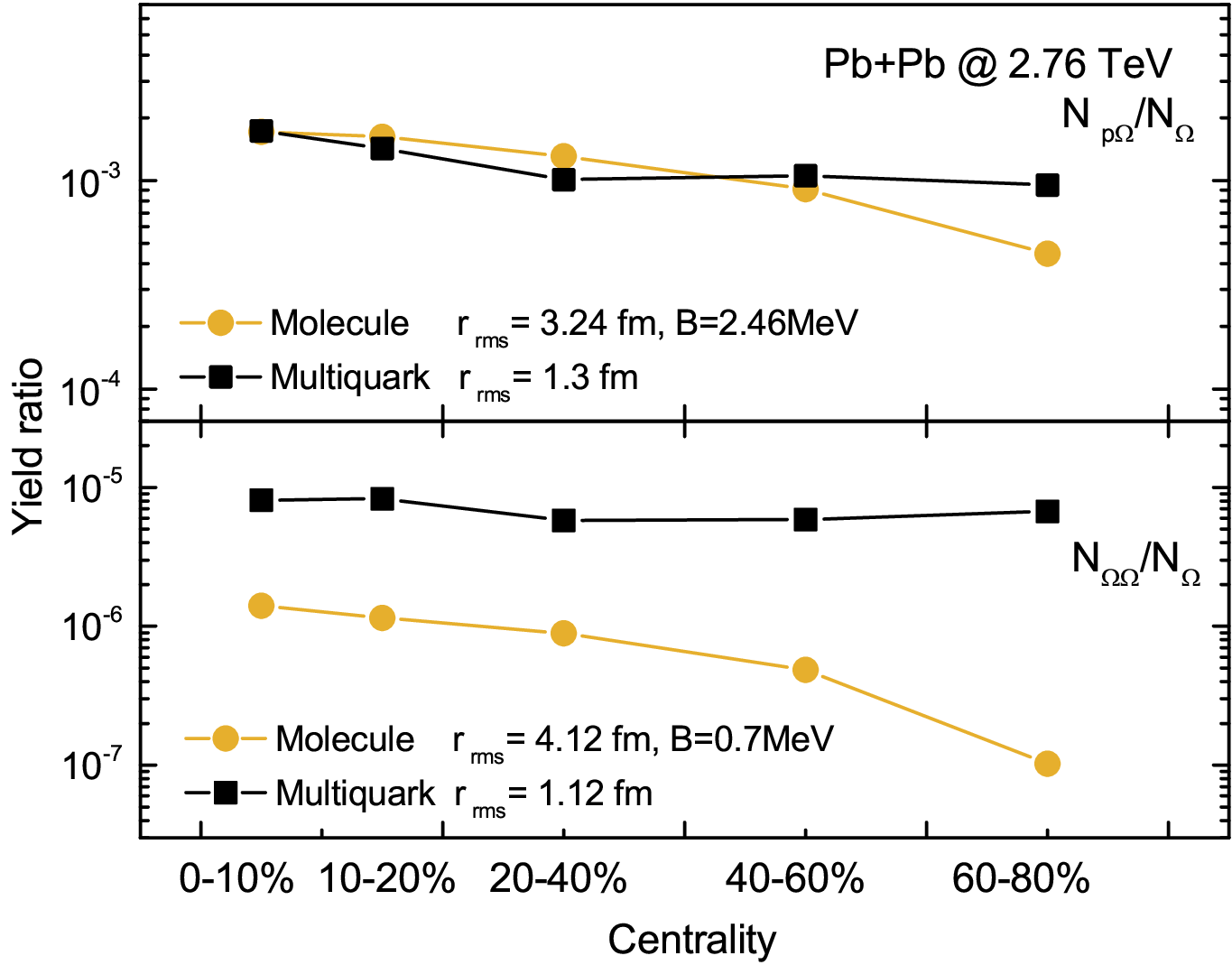}
\caption{\protect\small  Same with Fig. \ref{Figradio-test} but in the predicted root-mean-square radii. }
\label{Figradio-B}
\end{figure}
Fig.~\ref{Figradio-B} displays the yield ratios of $N_{p\Omega}/N_{\Omega}$ (top) and $N_{\Omega\Omega}/N_{\Omega}$ (bottom) with the estimated root-mean-square radii, i.e., $r_{\rm rms} = 1.3$ fm for multi quark coalescence of $p\Omega$ ($r_{\rm rms} = 1.12$ fm for $\Omega\Omega$), and $r_{\rm rms} = 3.24$ fm for hadron coalescence of $p\Omega$ ($r_{\rm rms} = 4.12$ fm for $\Omega\Omega$). It is seen that the $N_{p\Omega}/N_{\Omega}$ and $N_{\Omega\Omega}/N_{\Omega}$ drops rapidly in hadron coalescence but remains same trend in multi-quark coalescence as the centrality increasing. So the obviously different centrality dependence of the yield ratio for $p\Omega$ and $\Omega\Omega$ is a unique way to distinguish whether a structure is molecular or multi-quark state.  The difference in yield ratio between molecular and multi-quark state increases with the greater difference in root-mean-square radii.  

Although final-state scattering effects on the yields of  $p\Omega$ and $\Omega\Omega$  are absent if they are molecular states because they are produced after the kinetic freezeout of relativistic heavy-ion collisions, they may not be negligible in the scenario that they are multi-quark states produced at the chemical freezeout temperature, i.e., the beginning of the hadronic evolution. For example, hadronic effects on the yield of $D_{sJ}$(2317) and X(3875) as a multi-quark states produced at the chemical freezeout of relativistic heavy-ion collisions have been studied in Refs.~\cite{Chen:2007zp,Cho:2013rpa,Abreu:2016qci,Wu:2020zbx}. Although the hadronic final state interactions of multi-strange hadrons are generally thought to be weak~\cite{Shor:1984ui}, it will be interesting to investigate quantitatively this effects on the production of $\Omega\Omega$ and $p\Omega$, which requires detailed information about the interactions between the hadronic medium and these exotic states and is, however, currently unavailable. 

\section{conclusion}
In the present study, we have analyzed the production of $p\Omega$ and $\Omega\Omega$ in central Pb+Pb collisions at $\sqrt{s_{NN}} = 2.76$ TeV, employing a covariant coalescence model with blast-wave-like parametrization for constituent particle phase-space configurations at freeze-out. We have considered two scenarios: $p\Omega$ and $\Omega\Omega$ as either molecular or compact six-quark states. For the former, the phase-space configurations of constituent particles are determined from light nuclei and hypernuclei spectra; for the latter, they are determined from experimental data on hadrons such as $p$, $\Lambda$, $\phi$, $\Xi$, $\Omega^-$.

Our results indicate that in 0-10\% centrality, $p\Omega$ yields range from $6.41\times 10^{-4}$ to $1.32 \times 10^{-3}$ in the hadron coalescence and $3.89 \times 10^{-5}$ to $1.54 \times 10^{-3}$ in the quark coalescence, with $r_{\rm rms}$ spanning 0.5 to 5.0 fm. $\Omega\Omega$ yields vary from $6.25\times10^{-7}$ to $1.56\times10^{-6}$ in the hadron coalescence and $2.49\times10^{-7}$ to $8.80\times10^{-6}$ in the quark coalescence.   The STAR Collaboration's measurement of $^4$He yields ($8.6\times10^{-9}$~\cite{STAR:2011eej}) in $\sqrt{s_{NN}}=$200 GeV central Au+Au collisions suggests the feasibility of experimentally measuring $p\Omega$ and $\Omega\Omega$ yields.
Estimated root-mean-square radii for $p\Omega$ are 3.24 fm (hadron coalescence) and 1.30 fm (multi-quark coalescence), with yields of $9.92\times10^{-4}$ and $6.77\times10^{-4}$, respectively. For $\Omega\Omega$, these radii are 4.12 fm (hadron coalescence) and 1.12 fm (multi-quark coalescence), with yields of $8.13\times10^{-7}$ and $3.45\times10^{-6}$. Hadron coalescence predicts $p\Omega$ yields 1.46 times higher than multi-quark coalescence, and $\Omega\Omega$ yields in quark coalescence are 4.24 times those from hadron coalescence.

The $p\Omega$ and $\Omega\Omega$ yields in different centralities have been calculated by using the corresponding collision parameters. The yield ratios of $N_{p\Omega}/N_{\Omega}$ and $N_{\Omega\Omega}/N_{\Omega}$ going from central to peripheral collisions decrease for the hadron coalescence but changes mildly for the multi-quark coalescence. This distinct centrality dependence of yield ratios for molecular  and multi-quark states offers a promising way for distinguishing these two states. Our results thus suggest that measurements on the production of $p\Omega$ and $\Omega\Omega$ in relativistic heavy-ion collisions can  shed light on their internal structures and advance our understanding of the nature of the strong interaction.

\begin{acknowledgments}
We thank Kai-Xuan Cheng and Yu-Ting Wang for reading the manuscript.
This work was supported by the National Key Research and Development Project of China under Grant No.~2022YFA1602303, the National Natural Science Foundation of China under Grant No.~12105079, No.~12235010, No.~12375121, and the National SKA
Program of China No.~2020SKA0120300. 
\end{acknowledgments}
\bibliographystyle{elsarticle-num}

\end{document}